\documentstyle[preprint,prc,floats,aps]{revtex}
\tighten
\input BoxedEPS
\SetRokickiEPSFSpecial  
\HideDisplacementBoxes  

 \def\pp{^{\mathstrut}}  

\def\be{\begin{equation}}
\def\ee{\end{equation}}
\def\bea{\begin{eqnarray}}
\def\eea{\end{eqnarray}}

\begin{document}

\title{Molecular and Exotic Dibaryons and Other Hadrons\thanks{MIT-CTP \#2595. Invited
paper given at ISHEPP, Dubna, September 1996.}}

\author{E.L. Lomon}
\address{Center for Theoretical Physics and Laboratory for Nuclear Science\\ 
Massachusetts Institute of Technology\\
Cambridge, MA 02139\\
E-mail: lomon@mitlns.mit.edu}

\maketitle

\begin{abstract}Experimental evidence has been growing for the existence of both molecular
and exotic dibaryons. The former are dominated by hadron and the latter by quark-gluon
degrees of freedom.  Exotic dihadrons are of particular
interest because their properties are closely related to the parameters and
non-perturbative properties of QCD models. Molecular structures  provide information
about confinement. Purely exotic models, such as bag and valence quark models, ignore
the multi-hadron phase of confinement. Multi-configuration quark models and R-matrix
boundary condition models better represent the medium range repulsive effects leading
to confinement, predicting higher excitation energies of the exotic resonances.  The  $^1\!S_0$
exotic predicted at 2.70 GeV mass has been identified in pp spin observables, implying 
that there are several more even and odd parity exotics in the NN system
between 2.6 and 3.0 GeV\null. The method also predicts exotic strange and doubly strange
dibaryons (including a di-lambda unbound by about 0.13 GeV) and exotics in baryon and
meson channels. The R-matrix formalism also describes the properties of molecular structures
and enables a classification of these into several types. Observed and predicted examples of
each type are discussed, including the recently observed $d'$. 
\end{abstract}


\section{Di-Hadron Structures}

Just as nuclear bound states and resonances have been important in understanding the
nature of nuclear forces, the SU(3) structure of the single-hadron ($q\bar q$ and $q^3$) 
spectrum led to the quark model and QCD\null. It is  likely that the multi-hadron
spectrum can further reveal the properties of QCD as well as of the effective hadron
exchange forces. In particular the transition from a multi-hadron resonance to 
single hadrons is also a transition between the regions of perturbative asymptotic
freedom and non-perturbative confinement in QCD, an important process of which we
know little. 

But multi-hadron structures have many sources. Some are dominated by long-range forces
that can be described by hadron exchange potentials (confined quark clusters) and others
by the short-range quark and gluon degrees of freedom. 
The former are usually called molecular and the latter exotic structures. 
The long-range dominated
structures may arise from single- or multi-channel effects. We need to  classify
the structures by their experimental properties in order to deduce the implications for the
dynamics. We can learn about these relations from a model that adequately includes the
physics of both the long- and short-range regions. RGM calculations with a sufficiently large
number of quark configurations approach this behavior and have given useful results for
specific cases \cite{ref:aa}. The R-matrix method \cite{ref:bb}, which connects the short-range
perturbative QCD and the long-range hadronic region by a specified boundary condition,
has the advantage of a more adequate treatment of the non-perturbative region, simpler
calculations, and an analytic formulation that enables us to classify and describe
 structures according to the dynamical origin.

\section{Quark--Hadron Hybrid Model by R-Matrix Method}
\begin{figure}[ht]
\centering\leavevmode
\BoxedEPSF{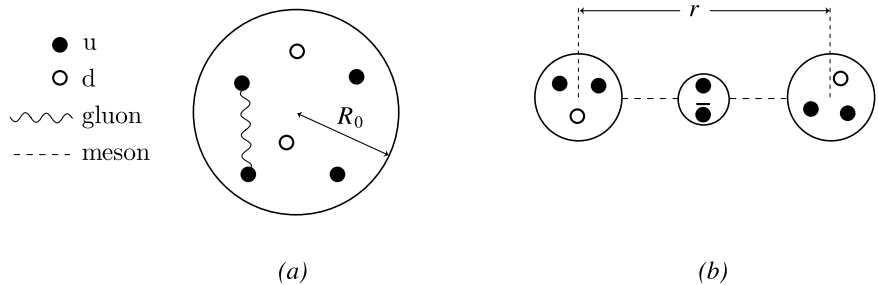 scaled 800}
\bigskip
\caption{Two protons at ({\it a\/}) small and ({\it b\/}) large separations.}
\label{fig:1}
\end{figure}
Figure~\ref{fig:1} represents two components of the pp wave function,  (a)~being the 
perturbative Dirac quark component with gluon exchange, and component~(b) described by a
two-proton wave function interacting via meson exchange. There are other components
like~(b) in which one or both nucleons are substituted for by the even and odd parity nucleon
isobars. Given a complete set of internal wave functions $\psi_i^c$ that vanish at
$R_0$, with total energy eigenvalues
$(W_i)$, R-matrix theory determines the external hadron wave function at energy $W$,
$\psi_W\pp (r)$,\footnote{The $\psi(r)\equiv r R(r)$, where $R$ is the radial wave function.}
 by the boundary condition at $r_0$ (corresponding to $R_0$)
\be
 r_0\psi^{\mathstrut\prime}_W  (r_0) = f(W) \psi_W\pp  (r_0)\quad\mbox{with}\quad
f(W) = f_0 +
\sum_i\frac{\rho_i}{W - W_i} \label{eq:1}
\ee
[the $\psi_W\pp $, $f(W)$, $f_0$, and $\rho_i$ are matrices that couple all isobar components
for a given $J$, $I$, and parity] and
\be
 \rho_i^{\alpha\beta} = -r_0\frac{d W_i}{d r_0} \xi_\alpha^i \xi_\beta^i \label{eq:2}
\ee
where the $\xi_\alpha^i$ are the fractional parentage coefficients of the $\alpha$ channel
component of $\psi_W\pp $ in the $\psi_i^c$ wave function. Integrating the Schr\"odinger
equation from $r_0$ to large~$r$ one obtains the S-matrix. The positions of the poles and
branch points of the S-matrix determine the structures (and any bound states) of the
two-hadron system.  The above model enables us to describe the effect of all dynamical
components on these singularities, with  implications for their experimental
characteristics. These features are all present in the simple case of two coupled channels with
different thresholds (such as
$NN \leftrightarrow \Delta\Delta$), with the neglect of the off-diagonal potential. Then the
scattering of an incoming channel of orbital angular momentum $L$ is described
by \cite{ref:cc}
\be
 S_L\pp \equiv \eta_L\pp e^{2i\delta_L} = \frac{{\cal J}^-_L (k, r_0)}{{\cal J}^+_L (k, r_0)}
\frac{f_{\rm eff}(W)- \theta_L (k, r_0)}{f_{\rm eff}(W) + \theta_L (k, r_0)} 
\label{eq:3}
 \ee
where 
\be
   f_{\rm eff}(W) = f_{L,L}(W) - \frac{\bigl(f_{L',L}(W)\bigr)^2}{f_{L',L'}(W) +
\theta_{L'}(k',r_0)}.
\label{eq:4}
 \ee
Where $k$ and $k'$ are the relativistic relative momenta in the $L$ and $L'$ channels,
\be
 \theta_L (k,r_0) = r_0 \frac{{\cal J}^{+\prime}_L (k, r_0)}{{\cal J}^+_L (k, r_0)}
\label{eq:5}
 \ee
and the ${\cal J}^{+,-}_L$ are the outgoing and incoming Jost functions, approaching the
spherical Hankel or the Coulomb functions at large range.
Their analytic properties determine that the $\theta_L$ are real for imaginary $k$ (below
threshold) and are complex for real~$k$. For a negligible strength potential
\be
\theta_L(k, r_0) = \frac{h^{(1)\prime}_L (k_0 r_0)}{h^{(1)}_L(k r_0)} 
\mathrel{\mathop{\hbox to 3em{\rightarrowfill}}\limits^{k\to 0}}
L + ik r_0.
\label{eq:5a}
 \ee

\section{Sources of Scattering Amplitude Structures}
 Poles of the S-matrix in the complex energy ($W$) plane will produce significant structure in
the energy dependence of scattering amplitudes if they are close enough to the physical cut.
Or, if they are on the real $W$ axis, then they must be below the minimum strongly coupled 
two-particle mass ($k$ imaginary for all channels) and represent a bound state. Also
significant structure may be caused by inelastic branch points ($k'=0$ for a coupled
channel) with a sufficiently strong discontinuity, representing strong coupling between
channels. 

\begin{figure}[ht]
\centering\leavevmode
\BoxedEPSF{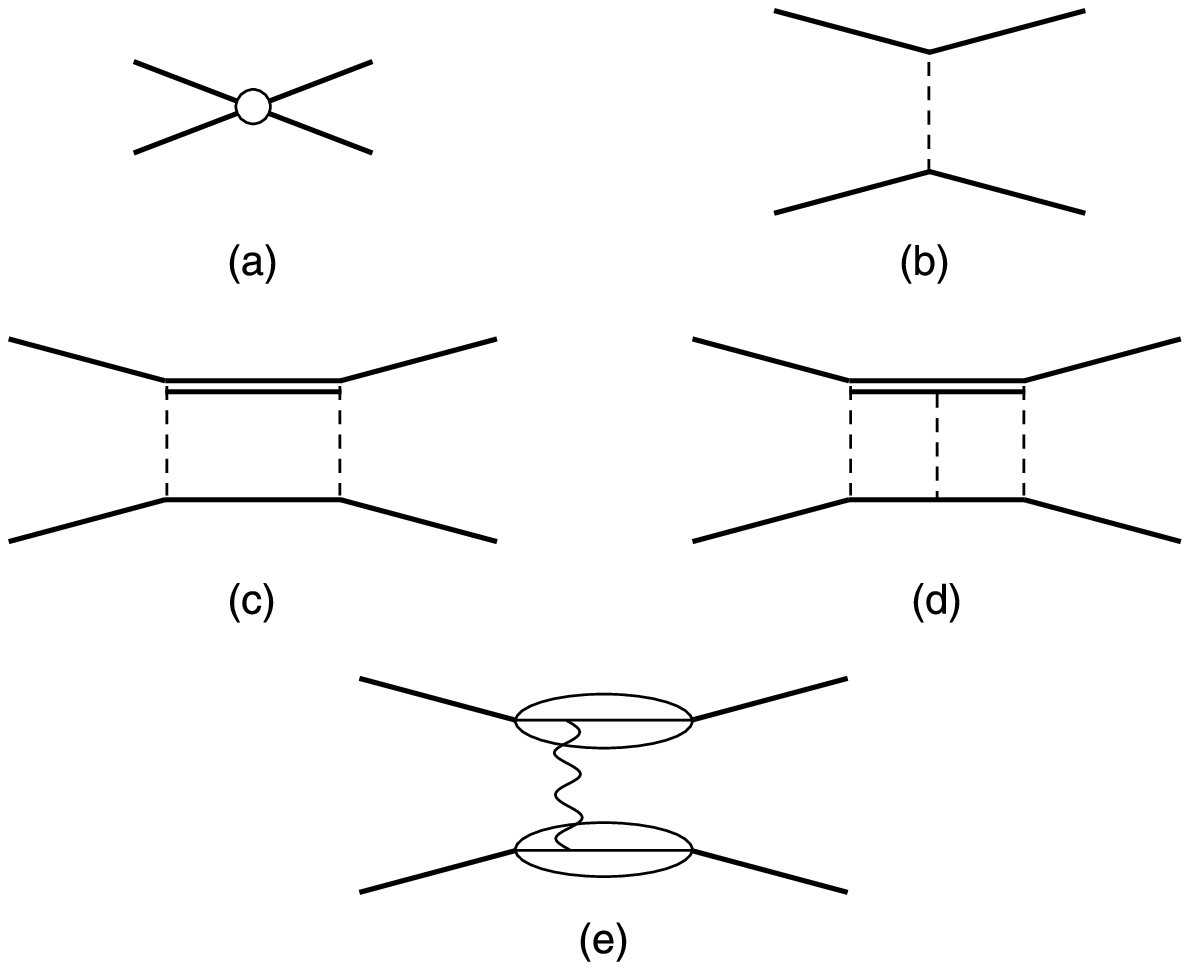 scaled 800}
\bigskip
\caption{Lowest order diagrams representing the various mechanisms that can cause structure
(resonances or bound states) in two-hadron interactions. For definiteness, baryons
(\raise.5ex\hbox{\vrule height 1pt width 10pt depth 0pt}) are shown  interacting via
mesons (\raise.5ex\hbox{\vrule height .5pt width 3pt depth 0pt}\,\raise.5ex\hbox{\vrule
height .5pt width 3pt depth 0pt}\,\raise.5ex\hbox{\vrule height
.5pt width 3pt depth 0pt}) or a core ($\circ$) representing the average effect of
short-range meson and gluon exchange, and intermediate states with baryon isobars 
(\rlap{\raise.75ex\hbox{\vrule height 1pt width 10pt depth 0pt}}\raise.25ex\hbox{\vrule
height 1pt width 10pt depth 0pt}) or quarks (\raise.5ex\hbox{\vrule height .5pt width 10pt
depth 0pt}) and gluons(\protect\BoxedEPSF{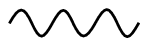 scaled 500}).}
\label{fig:2}
\end{figure}

The sources and character of each type of structure can be inferred from
Eqs.\,\ref{eq:1}--\ref{eq:4}. A pole of the S-matrix requires a zero of $\mbox{DEN}(W) \equiv
f_{\rm eff}(W) + \theta_L(k, r_0)$, and an observable structure in the continuum further
requires that the  $W$  dependence of DEN be sufficiently rapid in the vicinity of the zero.
The diagonal potential in channel~1 is represented by $\theta_L(k, r_0)$.  Eq.\,\ref{eq:5a} 
shows that if $f_{\rm eff}(W) < -L$ (see Fig.\,\ref{fig:2}(a)) there will be a bound state in the
absence of a potential. The attraction is then due either to the core
$f_{LL}$ or to the coupled channels in $f_{\rm eff}$. If the potential is attractive (see
Fig.\,\ref{fig:2}(b)),
$\theta_L - L$
 will be negative below threshold and there may be a bound state for $f_{\rm eff} >
-L$ or even for a repulsive core $f_{\rm eff} > L+1$. From Eq.\,\ref{eq:5} one can infer that in
the continuum (real
$k$), the pole will not be near the physical cut (producing a narrow resonance) unless $k$ is
small (an ``anti-resonance") or the potential is nonmonotonic producing a potential barrier
and rapid variation of $\theta_L(k,r_0)$. Alternatively the rapid variation may be in the
$f_{\rm eff}$ term, which by Eq.\ref{eq:4} requires either that the coupling $\left|
f_{L'L}\right|$ be large  (a large potential coupling will have the same effect -- see
Fig.\,\ref{fig:2}(c)), or that the denominator 
$\mbox{DEN}'=f_{L',L'}(w) + \theta_{L'}(k',r_0)$ vanish (see
Fig.\,\ref{fig:2}(d)) for ${k'}^2 < 0$. The
former causes a resonance below the second channel  threshold or even a bound state due
entirely to interchannel coupling \cite{ref:cc}. 
For small $\left|f_{LL'}\right|$ the associated pole \cite{ref:cc} is far from the physical cut and
the structure is an $L$-th  order cusp produced by the inelastic branch point. As
$\left|f_{LL'}\right|$ increases the pole comes nearer, and the structure forms into a broad
resonance below the threshold going further down in energy and becoming narrower with
further increase of $\left|f_{LL'}\right|$. 
The latter case (vanishing $\mbox{DEN}'$) is a state that would be bound if
there were no coupling to the first channel, but leaks into that channel through the coupling.
The smaller the coupling the narrower the resultant ``Dalitz--Tuan" resonance \cite{ref:dd}. 
The position of the resonance stays near the zero of $\mbox{DEN}'$. 
If the channels are only coupled by weak or electromagnetic forces, the result will be a very
narrow resonance. None of the above structures are importantly affected by the positions or
residues of the poles due to the interior quark states. 

However, in the vicinity of quark configuration energies (see Figs.\,\ref{fig:1} and
\ref{fig:2}(e)) the
$f$-matrix poles cause a rapid variation of $f_{\rm eff}$. 
The positions of these resonances are near the quark state energies, and the splittings for a
given quark configuration are determined by the one-gluon exchange matrix elements.  The
width of the resultant structure depends on $\rho^{\alpha\beta}_i$ (Eq.\,\ref{eq:1}), which is
decreased from the scale of $r_0^{-1}$ by the fractional parentage coefficients. This results in
widths of
$\sim 50$~MeV to $\sim 100$~MeV, easily detected with the resolution of multi-GeV beams.
It is important to note that any very narrow resonances observed experimentally are more
readily described by molecular ``Dalitz--Tuan" resonances than by exotic resonances induced
by quark degrees of freedom.

\section{Hadronic Examples of Various Structures}
There are many examples of potential bound molecular states, the deuteron and other nuclei
among them. Barrier potential resonance examples are represented by $\alpha$ unstable and
fissioning nuclei. The $^1D_2$ and $^3F_3\, NN$ states near the $N\Delta$ threshold are
well understood \cite{ref:bb,ref:ee} as driven by the $N\Delta$ threshold branch point.
The
$Y^*_0(1405)$ and the
$S^*$ are caused by the coupling to $K^- p$ and $K\bar K$ systems, respectively; either
through the strong coupling  or by the ``Dalitz--Tuan" formation of a quasi-bound
state \cite{ref:dd}. The recently observed \cite{ref:ff} $d'(2060)$ has a width $\leq 1$~MeV,
and could be described \cite{ref:ff,ref:gg} as a bound state of a nucleon and an odd-parity
nuclear isobar
$I(J^\rho) = O(O^-)$ coupled electromagnetically to an $NN\pi$ state.

\begin{figure}[ht]
\centering\leavevmode
\BoxedEPSF{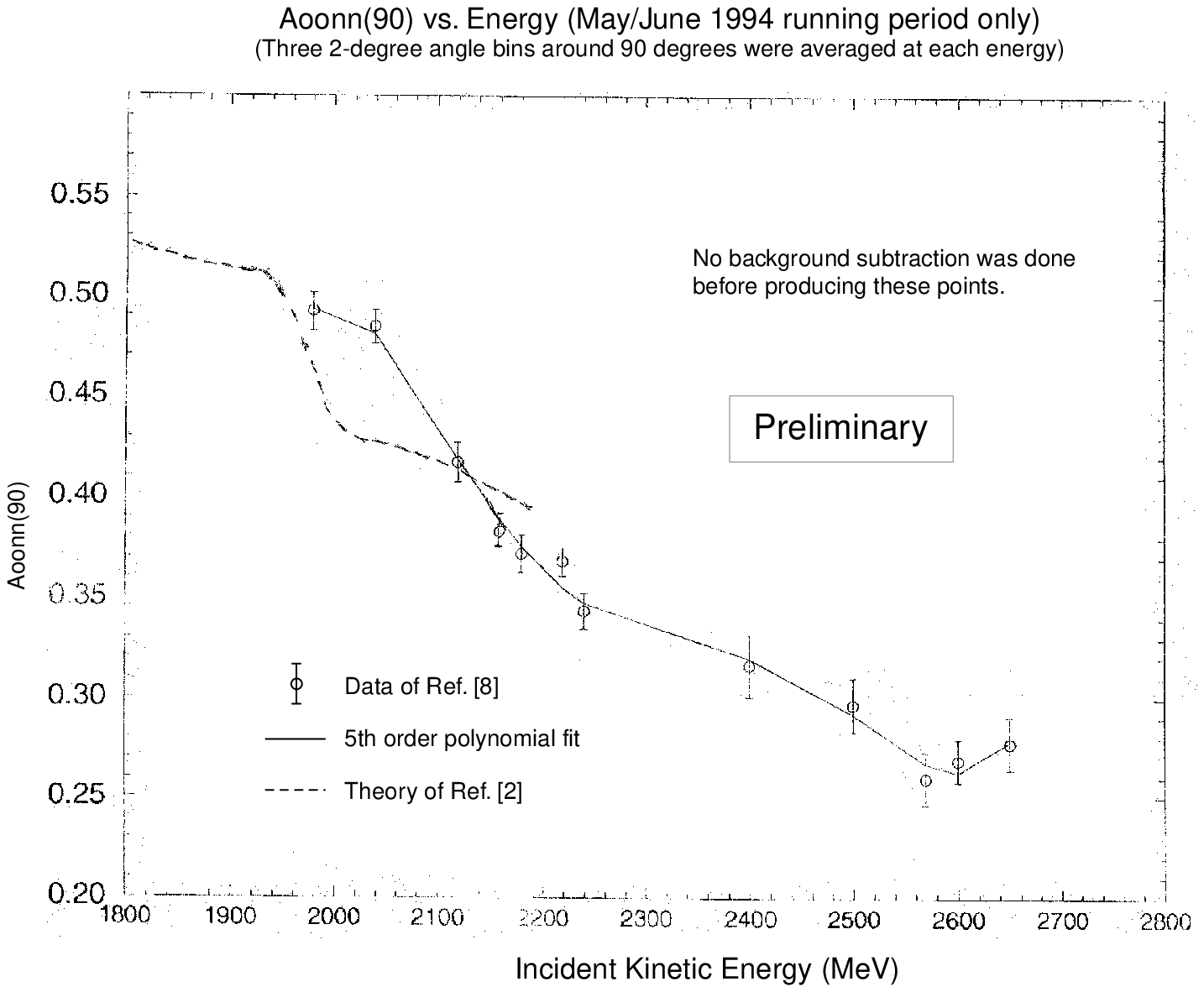}
\bigskip
\caption{}
\label{fig:3}
\end{figure}

The spectra
of the quark exotic $[q(1s_{\frac12})]^6$, $[q(1s_{\frac12})]^5 q(1p_{\frac12})$, 
$[q(1s_{\frac12})]^5 s(1s_{\frac12})$,\linebreak
 $[q(1s_{\frac12})]^4 [s(1s_{\frac12})]^2$, and
$[q(1s_{\frac12})]^4 \bar s(1s_{\frac12}) $ quark configurations have been predicted by the
R-matrix method \cite{ref:bb,ref:gg,ref:hh} ($q$ represents $u$ or $d$ quarks). Constituent
quark calculations predict similar spectra, at about 100~MeV less mass. The lowest mass
$I=1\,NN$ state is predicted at 2.7~GeV$/c^2$ mass. Several experiments have observed
structure at this energy, the most significant of which (see Fig.\,\ref{fig:3}) is the energy
dependence of the spin correlation
$Ann(pp)$ \cite{ref:hh}. The prediction made in 1987 \cite{ref:bb} is in good agreement with
the data. Some recent
$np$ data is consistent with the lowest predicted $I=0$ state at 2.63~MeV \cite{ref:ii},
but data is required in finer energy steps. We await further experiments in the $NN$,
$\Lambda N -
\Sigma N$, $\Lambda\Lambda - \Xi N$, and $KN$ systems. 

\section*{Acknowledgments}
 This work is supported in part by funds
provided by  the U.S.~Department of Energy (D.O.E.) under contract
\#DE-FC02-94ER40818.

\def\Journal#1#2#3#4{{#1} {\bf #2}, #3 (#4)}

\def\AP{{\em Ann. Phys.} (N.Y.)}
\def\NCA{\em Nuovo Cimento}
\def\NIM{\em Nucl. Instrum. Methods}
\def\NIMA{{\em Nucl. Instrum. Methods} A}
\def\NPB{{\em Nucl. Phys.} B}
\def\PLB{{\em Phys. Lett.}  B}
\def\PRL{\em Phys. Rev. Lett.}
\def\PRD{{\em Phys. Rev.} D}
\def\PPNP{\em Prog. Part.  Nucl. Phys.}
\def\SOVJ{\em Sov. J. Nucl. Phys.}
\def\ZPC{{\em Z. Phys.} C}

 \end{document}